\documentstyle[12pt]{article}

\begin{document}

\baselineskip=14pt plus 0.2pt minus 0.2pt
\lineskip=14pt plus 0.2pt minus 0.2pt

%***********************
\newcommand{\beq}{\begin{equation}}
\newcommand{\eeq}{\end{equation}}
\newcommand{\bea}{\begin{eqnarray}}
\newcommand{\eea}{\end{eqnarray}}
\newcommand{\da}{\dagger}
\newcommand{\dg}[1]{\mbox{${#1}^{\dagger}$}}
\newcommand{\hlf}{\mbox{$1\over2$}}

%**********************

\begin{flushright}
quant-ph/9506008 \\
 LA-UR-95-1772 \\
\end{flushright}

\begin{center}
\large{\bf
Arbitrary-Order Hermite  Generating Functions \\
for Coherent and Squeezed States}

\vspace{0.25in}

\large
\bigskip

Michael Martin Nieto\footnote{\noindent  Email: mmn@pion.lanl.gov}\\
{\it Theoretical Division, Los Alamos National Laboratory\\
University of California\\
Los Alamos, New Mexico 87545, U.S.A. \\}

\vspace{0.25in}

 D. Rodney Truax\footnote{Email:  truax@acs.ucalgary.ca}\\
{\it Department of Chemistry\\
 University of Calgary\\
Calgary, Alberta T2N 1N4, Canada\\}

\normalsize

\vspace{0.3in}

{ABSTRACT}

\end{center}
\begin{quotation}

%****************************
\baselineskip=0.33In
%***************************

For use in calculating higher-order coherent- and squeezed-
state quantities,
we derive  generalized generating functions  for the Hermite
polynomials.  They are given by
$\sum_{n=0}^{\infty}z^{jn+k}H_{jn+k}(x)/(jn+k)!$,
for arbitrary  integers $j\geq 1$ and
$k\geq 0$.  Along the way, the sums with the Hermite polynomials replaced by
unity are also obtained.  We also evaluate the action of the
operators $\exp[a^j(d/dx)^j]$ on well-behaved functions and apply them to
obtain other sums.

\vspace{0.25in}

%\noindent PACS: 02.20.Gp, 02.30.Lt

\end{quotation}

\newpage

%********************************************************************
\baselineskip=.33in
%******************************************************************

\section{Introduction}

In calculations involving coherent and squeezed states, infinite
sums involving Hermite polynomials are extremely useful.
Even so, knowledge of such sums is limited.  For example,
if we define the generalized generating functions for the Hermite
polynomials as
\beq
G(j,k) = \sum_{n=0}^{\infty}\frac{z^{jn+k}H_{jn+k}(x)}{(jn+k)!}~,
    \label{gkj}
\eeq
for arbitrary integers $j\geq 1$ and $k\geq 0$,
it is only the specific  generating functions
\bea
G(1,0) &=& \exp[-z^2+2xz]~,  \label{g1} \\
G(2,0) &=& \exp[-z^2] \cosh(2xz)~, \label{g2} \\
G(2,1) &=& \exp[-z^2] \sinh(2xz)~, \label{g21}
\eea
that are well-known  \cite{mos}.
Physically,
Eqs. (\ref{g1})-(\ref{g21}) are very important as they describe the
decomposition of ordinary, even, and odd  coherent states, respectively,
into  the harmonic-oscillator number states \cite{nt}.

Motivated by this situation, in this paper we will show that  the general
generating functions \cite{wilf}  $G(j,k)$
are given by
\beq
G(j,k)= \frac{1}{j}\sum_{l=1}^{j}
  \frac{\exp[-z^2 e^{i4\pi l/j}]\exp[2xz e^{i2\pi l/j}]}
       {[e^{i2\pi l/j}]^k}~.  \label{gans}
\eeq
Along the way we will also obtain the sums
\beq
S(j,k)=\sum_{n=0}^{\infty}\frac{z^{jn+k}}{(jn+k)!}
= \frac{1}{j}\sum_{l=1}^{j}
  \frac{\exp[z e^{i2\pi l/j}]}
       {[e^{i2\pi l/j}]^k}~. \label{skj}
\eeq

We will use the same techniques to evaluate
\beq
I_j = \exp\left[\left(a\frac{d}{dx}\right)^j\right]g(x)~,
\eeq
and apply it to obtain further sums involving Hermite polynomials.

%******************************************************************

\section{The Sums $S(j,k)$}

First note the special cases
\bea
S(1,0) &=& e^z~,  \label{s1}   \\
S(2,0) &=& \cosh z = \frac{1}{2}\left[e^z+e^{-z}\right]~.
     \label{s2}
\eea
A few other sums are known.  In particular, $S(3,0)$ can be obtained
from the known formula for $S(3,1)$: \cite{hansen}
\beq
S(3,0)=\left(\frac{d}{dz}\right)S(3,1)
        =\left(\frac{d}{dz}\right)
         \left[\frac{1}{3}e^z-\frac{2}{3}e^{-z/2}
   \cos\left(3^{1/2}\frac{z}{2} +\frac{\pi}{3}\right)\right]~.
\eeq
With a little algebra, the result can be written in the form
\beq
S(3,0)=\frac{1}{3}\left(\exp[ze^{i2\pi/3}]+\exp[ze^{i4\pi/3}]
          +\exp[z]\right)~.  \label{s3}
\eeq

But from Eqs. (\ref{s1}), (\ref{s2}), and {\ref{s3}) we can now guess what
the answer for $S(j,0)$ is.  It is
\beq
S(j,0)= \frac{1}{j}\sum_{l=1}^{j} \exp[z e^{i2\pi l/j}]~.
\eeq
The phase factors in the exponentials are the $j$th roots of unity.
Therefore, in the sum of the  Taylor series expansions of the exponentials,
after the $z^0$ term,  the terms proportional to
$z$, $z^2$, $\dots$, $z^{j-1}$
cancel to zero. The next  non-zero term is the $z^j$
term.  Similarly for the higher-order terms.

This argument can be made precise by using the geometric series
\beq
\sum_{l=1}^j r^l = \frac{r(r^j-1)}{r-1}~.
\eeq
Taking $r=\exp[i2\pi/j]$, $r=\exp[i4\pi/j]$, $r=\exp[i6\pi/j]$, $\dots$,
gives the result.

Then, $S(j,k)$ of Eq.  (\ref{skj})  is obtained by observing that
\beq
S(j,k) = \left(\int dz \right)^k S(j,0)~.
\eeq

%******************************************************************

\section{The Generating Functions $G(j,k)$}

Now we are ready to calculate the generating functions $G(j,k)$.
In Eq. (\ref{gkj}) for $G(j,k)$, substitute the definition
\cite{mos}
\beq
H_n(x) = (-1)^n\exp[x^2]\frac{d^n}{dx^n}\exp[-x^2]~.
\eeq
One then has
\beq
G(j,k) = \exp[x^2]\left[ \sum_{n=0}^{\infty}
     \frac{(-z\frac{d}{dx})^{jn+k}}{(jn+k)!}\right] \exp[-x^2]~,
\eeq
which from Eq. (\ref{skj}) is equal to
\beq
G(j,k) =  \exp[x^2] \left[ \frac{1}{j}\sum_{l=1}^{j}
  \frac{\exp[-z e^{i2\pi l/j} \frac{d}{dx}]}
       {[e^{i2\pi l/j}]^k} \right] \exp[-x^2]~.
\eeq
But now using the relation (see below)
\beq
\exp\left[a\frac{d}{dx}\right]~g(x) = g(x+a)~,  \label{i1}
\eeq
one obtains the final result for $G(j,k)$ given in Eq. (\ref{gans}).

%*************************************

\section{The Operators $\exp[a^j(d/dx)^j]$}

The techniques we have used above can be applied to other situations.
For example, we may ask if the result of Eq. (\ref{i1}),
\beq
I_1  = \exp\left[a\frac{d}{dx}\right]~g(x)
=\sum_{n=0}^{\infty} \frac{a^ng^{[n]}(x)}{n!} = g(x+a)~,  \label{i11}
\eeq
can  be generalized  to $j>1$:
\beq
I_j = \exp\left[\left(a\frac{d}{dx}\right)^j\right]g(x)
=\sum_{n=0}^{\infty} \frac{a^{jn}g^{[jn]}(x)}{n!}~. \label{ij}
\eeq

First consider the special case $j=2$:
\beq
I_2 = \sum_{n=0}^{\infty} \frac{a^{2n}g^{[2n]}(x)}{n!}~.
\eeq
Using the doubling formula
\beq
\Gamma(n+1/2) = \pi^{1/2}\frac{(2n)!}{2^{2n}n!}~,
\eeq
one obtains
\bea
I_2&=&\frac{1}{\pi^{1/2}}\sum_{n=0}^{\infty} \frac{(2a)^{2n}}{(2n)!}
            \Gamma(n+1/2)g^{[2n]}(x) \nonumber \\
&=&\frac{1}{2\pi^{1/2}}\sum_{n=0}^{\infty} \frac{(2a)^{n}}{n!}
     \Gamma\left(\frac{n+1}{2}\right)[g^{[n]}(x)+(-1)^ng^{[n]}(x)]\nonumber \\
&=&\frac{1}{2\pi^{1/2}}\int_0^{\infty}\frac{ds~e^{-s}}{s^{1/2}}
\sum_{n=0}^{\infty} \frac{(2as^{1/2})^{n}}{n!}
            [g^{[n]}(x)+(-1)^ng^{[n]}(x)]\nonumber \\
&=&\frac{1}{2\pi^{1/2}}\int_0^{\infty}\frac{ds~e^{-s}}{s^{1/2}}
            [g(x+2as^{1/2})+g(x-2as^{1/2})]~.  \label{i2}
\eea
In the above, the use of $(-1)^n$ allowed us to cancel out the odd derivatives
as we changed the sum from second-order terms to first-order terms.  The use
of the integral representation of the $\Gamma$ function then led to the final
result.

Note that by the change of variables
\beq
s = {(y-x)^2}/{4a^2}~,
\eeq
our result can be put in the form of the known result  \cite{kato}
\beq
I_2=\frac{1}{(4a^2\pi)^{1/2}}\int_{-\infty}^{\infty}
   \exp\left[-\frac{(y-x)^2}{4a^2}\right]~g(y)dy~.
\eeq
(This formula will be useful in obtaining further sums in the next
section.)

The way to generalize to arbitrary $j\geq 2$ is now,
{\it in principle},  clear.
First one uses the multiplication formula
(which is the generalization of the doubling
formula)  in the form
\beq
\frac{1}{n!} = \frac{j^{jn+1/2}}{(2\pi)^{(j-1)/2}}
\frac{\prod_{k=1}^{j-1} \Gamma(n+k/j)}{(jn)!}~.
\eeq
Then, just as before,
the use of $\exp[i2\pi l/j]$ will allow the cancelation of  undesired terms
in the complete sum over $n$.  Here those terms are the
$g^{[p\neq nj]}$.  When  all is done,
the end result is
\bea
I_j &=& \frac{1}{(2\pi)^{(j-1)/2}j^{1/2}}
\int_0^{\infty}\frac{dx_1~e^{-x_1}}{x_1^{1/j}}
\int_0^{\infty}\frac{dx_2~e^{-x_2}}{x_2^{2/j}}\dots
\int_0^{\infty}\frac{dx_{j-1}~e^{-x_{j-1}}}{x_{j-1}^{(j-1)/j}}
\nonumber \\  &~&
\sum_{l=1}^{j}g\left(x+ja(x_1x_2\dots x_{j-1})^{1/j}e^{i2\pi l/j}\right)~.
\eea

This result, although in closed form, is usually too complicated to
yield results in terms of elementary functions.
For instance, even for $j=3$, changing variables to $x_i = y_i^3$
yields
\beq
I_3 = \frac{9}{(2\pi)3^{1/2}}
\int_0^{\infty} dy_1~y_1 e^{-y_1^3}
\int_0^{\infty} dy_2~e^{-y_2^3}
\sum_{l=1}^{3}g\left(x+3ay_1y_2 e^{i2\pi l/3}\right)~.
\eeq

But there is also
another point that must be mentioned.  In Ref. \cite{fisher}, it was
observed that naive higher-order squeeze operators of the form
$\exp[z_ja^{\dagger j} - z_j^* a^j]$, $j>2$, are not Gaussian-integrable
 operators.   Therefore, the convergence of the $I_j$,  $j>2$,
depends on the form of $g(x)$.

Finally, for $j>3$, Baker \cite{baker} has observed that using the Hankel
contour integral for $1/\Gamma$ will lead to a formally simpler answer.
Starting from Eq. (\ref{ij}), one has
\beq
I_j=\sum_{n=0}^{\infty} \sum_{l=1}^{j}
\frac{a^{n}e^{i2\pi ln/j}g^{[n]}(x)}{j~[(n/j)]!}~,
\eeq
where $[(n/j)]!$ is not clearly defined for $n$ not a multiple of $j$.
However,
because of the sum over $l$, only the $n$-terms which are multiples of
$j$ will sum to non-zero.  Therefore, $[(n/j)]!$ can be replaced by
$1/\Gamma(1+n/j)$, which in turn can be replaced by the Hankel contour
integral.  Unity in the form of $\Gamma(n+1)/n!$ can be inserted, and
$\Gamma(n+1)$ replaced by its integral representation. Then, a little
algebra yields
\beq
I_j = \frac{1}{j2\pi i}\int_C\frac{dt e^{-t}}{t}
      \int_0^{\infty}ds~se^{-s}
      \sum_{l=1}^{j} g\left(x+ast^{-1/j}e^{i\pi(2l-1)/j}\right)~.
\eeq
$C$ is the contour starting just above the $x$-axis at $(\infty, i\epsilon)$,
coming in and encircling the origin counterclockwise, going out to
$(\infty, -i\epsilon)$, and closing.
It is straightforward to demonstrate that $I_1$ and $I_2$ of Eqs.
(\ref{i11}) and (\ref{i2}) are special cases.

%******************************************************************

\section{Discussion}

With our methods, it is possible to obtain further quantities. Sums of the type
\beq
K(j,k,p,q) = \sum_{n=0}^{\infty}\frac{z^{jn+k}H_{jn+k}(x)}{(pn+q)!}
\eeq
become amenable to analysis, as well as even more complicated sums.

For example, one can immediately write
\bea
K(2,0,1,0)&=&\exp[x^2]\exp\left[\left(-z\frac{d}{dx}\right)^{2}\right]
           \exp[-x^2] \nonumber \\
     &=&\frac{\exp[x^2]}{(4z^2\pi)^{1/2}}\int_{-\infty}^{\infty}
   \exp\left[-\frac{(y-x)^2}{4z^2}\right]\exp[-y^2]dy
\nonumber  \\
  &=& \frac{1}{[1 + 4z^2]^{1/2}}\exp\left[\frac{4z^2x^2}{1+4z^2}\right]~.
\eea
This describes the squeezed ground state.  Also, one  has
\bea
K(2,1,1,0)& =&\exp[x^2]\left(-z\frac{d}{dx}\right)
      \left[K(2,0,1,0)\exp[-x^2]\right] \nonumber \\
     &=& \frac{2zx}{[1+4z^2]^{3/2}}
               \exp\left[\frac{4z^2x^2}{1+4z^2}\right]~.
\eea
Adding these two yields the known sum \cite{correct}
\beq
\sum_{n=0}^{\infty}\frac{z^{n}H_{n}(x)}{[[n/2]]!}
=\frac{1+2zx + 4z^2}{[1 + 4z^2]^{3/2}}
     \exp\left[\frac{4z^2x^2}{1+4z^2}\right]~,
\eeq
where in the above $[[\cdot]]$ denotes the greatest integer function.

%*****************

\section*{Acknowledgements}

We wish to thank George Baker, Jr.  and Jim Louck for
their very helpful  observations and
comments.  The work of
MMN and DRT was supported by the US Department of Energy and the Natural
Sciences and Engineering Research
Council of Canada, respectively

%******************************************************************

\hspace{2.in}


\begin{thebibliography}{99}
\bibitem{correct} On p. 253 of Ref. \cite{mos}, the factor
       $\sqrt{1+4z^2}$ should be cubed.  In Eq. (49.4.1) of Ref.
\cite{hansen}, the factor $\Gamma[(k/2)+1]$ should be replaced by
              $[[k/2]]$.
\end{thebibliography}
\end{document}